\documentclass[journal=jacsat,manuscript=article]{achemso}

\usepackage{graphicx}
\usepackage{dcolumn}
\usepackage{multirow}
\usepackage{bm}

\usepackage[utf8]{inputenc}
\usepackage[T1]{fontenc}
\usepackage{mathptmx}
\usepackage{etoolbox}
\usepackage{physics}
\usepackage{color} 
\usepackage{hyperref}
\usepackage{amsmath}
\usepackage{booktabs}       
\usepackage{threeparttable}
\usepackage{appendix}
\usepackage{ulem}
\usepackage{units}
\usepackage{soul} 

\usepackage[inline]{enumitem}
\usepackage[version=3]{mhchem} 
\newcommand{\im}{\textup{i}}
\author{Yong-Tao Ma}
\affiliation{%
Department of Chemistry, School of Science and Research Center for Industries of the Future, Westlake University, Hangzhou, Zhejiang 310024, China
}%

\author{Wenjie Dou}
\affiliation{%
Department of Chemistry, School of Science and Research Center for Industries of the Future, Westlake University, Hangzhou, Zhejiang 310024, China
}%
\alsoaffiliation{%
Institute of Natural Sciences, Westlake Institute for Advanced Study, Hangzhou, Zhejiang 310024, China
}%
\alsoaffiliation{%
Key Laboratory for Quantum Materials of Zhejiang Province, Department of Physics, School of Science and Research Center for Industries of the Future,
Westlake University, Hangzhou, Zhejiang 310024, China
}%
\email{douwenjie@westlake.edu.cn} 

\title[OSH-themostat]
  {Orbital Surface Hopping with an Electron Thermostat Yields Accurate Dynamics and Detailed Balance}

\abbreviations{IR,NMR,UV}
\keywords{American Chemical Society, \LaTeX}

\begin{document}

\begin{tocentry}

\includegraphics[height=4.6cm]{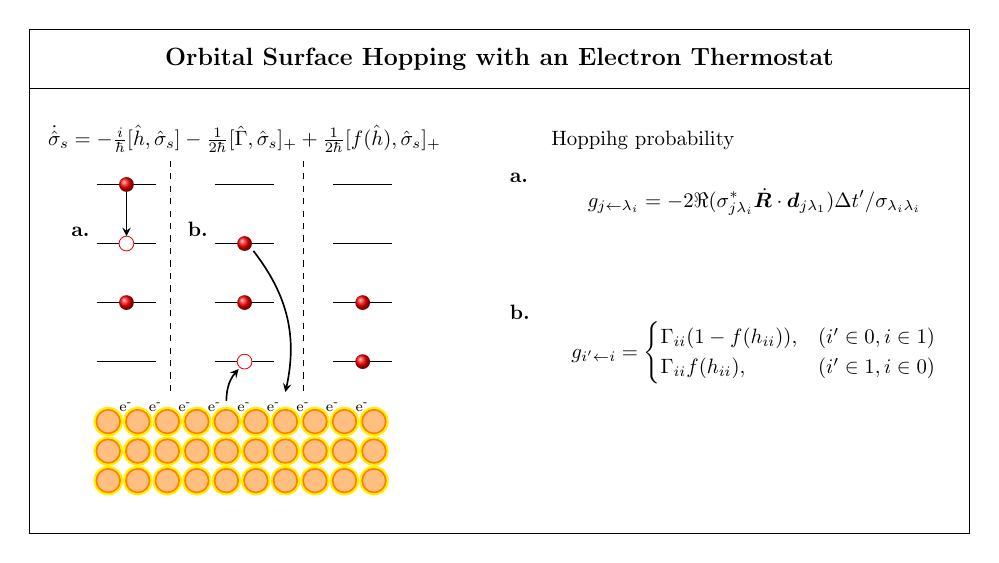}

\end{tocentry}

\begin{abstract}
In mixed quantum-classical simulations of molecule-metal surface interactions, the discretization of the metallic electronic continuum typically
results in a closed-system representation that fails to capture the open-system nature of the true physical process.
This approximation can introduce significant artifacts, including deviations in the dynamical evolution and a violation of the principle of detailed balance.
To address this fundamental challenge, we introduce an electronic thermostat into our previously developed orbital surface hopping (OSH) framework,
generalizing the method to efficiently handle many discrete electronic states. We first outline the derivation of electronic thermostat orbital surface hopping,
where the amplitude of the electronic thermostat is well justified. We then demonstrate that this method can reproduce accurate dynamics and detailed balance in long time,
whereas without electronic thermostat the detailed balance is violated. Thus, this method offers a reliable tool for studying nonadiabatic dynamics near metal surfaces. 

\end{abstract}
\maketitle 
\section{Introduction}
Understanding and modeling nonadiabatic effects in surface inelastic scattering and reaction dynamics represent
an area of increasing scientific interest.\cite{jiang-prl-2024,guohua-h2Au-2020,Wodtke20a,wodtke-2024-review,Rainer-interference-2025,sqiang-heom-Au-2025,thoss-heom-2025}
These effects are pivotal in governing key processes involving electron and energy transfer at surfaces \cite{tully2000rev,wodtke-tully2021,wodtke2004}. 
Accurate simulation of such phenomena requires theoretical frameworks capable of capturing nonadiabatic dynamics. For realistic systems,
however, the large number of degrees of freedom renders fully quantum mechanical methods, such as Multi-Configuration 
Time-Dependent Hartree (MCTDH)\cite{Beck20001} and Hierarchical Equations of Motion (HEOM)\cite{Qiang2019}, computationally prohibitive. This computational bottleneck has driven the development of mixed quantum-classical methods
as a practical alternative. Among these, the Ehrenfest and surface hopping approaches have gained widespread adoption due to
their conceptual clarity and computational efficiency. It is critical to note that, unlike their application in the gas phase,
both methods require specific adaptations to accurately capture the unique electronic characteristics of metal surfaces.

In the Electron Friction approach, nuclear motion is classically propagated on the ground-state potential energy surface
under the influence of an additional frictional force originating from metal-molecule interactions.\cite{joseph2017,james1993,Suhl1975,Newns1995,HGM1995,Joseph2015,Joseph2016,Felix2011,Daniel2012,PERSSON1980} 
In contrast, the Classical Master Equation (CME) and Broadened Classical Master Equation (BCME) frameworks\cite{QCMEalg-D2015,QCMEderive-D2015}, 
the electronic degrees of freedom of the metal are treated within an open quantum system framework. 
In these methods, the electronic states of the metal surface are partially traced out, and electron transfer between the metal
and the adsorbate is modeled as stochastic hops between molecular neutral and anionic states—with the latter broadened in the BCME formulation.
The metal-molecule interaction is incorporated through a generalized damping term. Notably, the BCME approach has shown remarkable accuracy
across a broad spectrum of coupling strengths, ranging from the weak to the strong coupling regime.\cite{miao18} However, a recognized limitation of the BCME method
is its restriction to systems with effectively one electronic state, preventing its direct application to problems involving multiple active electronic states.\cite{subonic-review-jpca-2020}

Perhaps the most prominent method in this category is Independent Electron Surface Hopping (IESH).\cite{tully2009s,tully2009j} 
In contrast to the BCME approach, IESH explicitly retains the electronic degrees of freedom of the metal.
Within this framework, nuclear dynamics evolve on adiabatic potential energy surfaces constructed
by coupling the molecular anionic state with discrete electron states of the metal. Transitions between these surfaces (“hops”)
are governed by probabilities derived from the time-dependent electronic wavefunctions.
The IESH method has accurately reproduced experimental results for NO scattering on the Au(111) surface,\cite{tully2009s} 
showing excellent agreement with measured vibrational energy relaxation and electron population dynamics.
Furthermore, the theoretical simulations revealed a steering mechanism that provides additional insight into the scattering process.
The conceptual foundation of IESH is closely related to the trajectory surface-hopping time-dependent Kohn-Sham (TDKS) approach
pioneered by Prezhdo and coworkers.\cite{oleg2005,oleg2006}
This lineage has inspired the development of several popular software packages, such as PYXAID\cite{pyxaid2013}, Hefei-NAMD\cite{hefei-namd-2019} and GTSH\cite{gtsh-2023},
which are designed for simulating nonadiabatic molecular dynamics in condensed matter systems.
In our recent work, we theoretically demonstrated that, within the independent electron approximation,
the hopping probabilities can be formulated directly in terms of orbital densities involved in each hopping event.\cite{dou-osh-2025}
This formulation, which we term the Orbital Surface Hopping (OSH) method, offers a more physically transparent picture
by tracking electron evolution through individual orbitals at the metal surface. The OSH approach achieves accuracy comparable to IESH
but at a substantially reduced computational cost.

A fundamental limitation of the original IESH and OSH methods lies in their discretization of the electronic continuum,
which enforces a closed-system representation. This approximation severs the essential energy dissipation pathway into the metal surface,
resulting in spurious conservation of the total energy. Such behavior is unphysical given the vast heat capacity of a metal
and introduces artifacts including erroneous long-time dynamics and a violation of detailed balance.
While enlarging the basis set of discrete states could better approximate the continuum, it does so at a prohibitive computational cost.
This inherent shortcoming was recognized by the Tully group
three years after IESH was first introduced, prompting them to propose an electronic thermostat methodology.\cite{tully2012fd} 
The approach was motivated by two physical insights: first, that electron–electron scattering occurs on a much shorter timescale than nuclear motion,
and second, that the probabilities for transferring a high-energy electron to a low-energy hole should obey a Boltzmann distribution.
Implemented in a Monte Carlo framework, this thermostat successfully recovers correct equilibrium populations and kinetic energy.\cite{miao18}
Despite its conceptual importance, the method initially received limited attention. Only recently has its utility been demonstrated in other contexts:
the Prezhdo group employed it in an Auger process to enforce electron number conservation\cite{prezhdo-auger-2025},
and the Jiang group used it in initial electron states sampling for NO scattering on Au(111).\cite{jiang-co-Au-jpcc-2025}
Building on these developments and grounded in open quantum system theory combined with the concept of orbital densities,
we have formulated a new electronic thermostat strategy. This methodology is implemented in our OSH approach.
We benchmark its performance against both the Tully's electronic thermostat and the reference HEOM method.
Results indicate that our proposed electron thermostat significantly improves the approach to equilibrium electronic populations and
yields more physically accurate kinetic energy distributions.   
 
 This paper is organized as follows. Section II details the computational methodology. The results and corresponding discussion are presented in Section III.
 Finally, Section IV provides the concluding remarks.

\section{Methodology}

In general, we define the Hamiltonian for a non-interacting system coupled to a continuous bath as:
\begin{equation}
\hat{H}_{total} = \hat{H}_{s} + \hat{H}_{b} + \hat{H}_{c},
\end{equation}
in which, the system Hamiltonian is:
\begin{equation}
\hat{H}_s = \sum_{mn} h_{mn}(\hat{R}) \hat{d}_m^\dagger \hat{d}_n + U(\hat{R}) + T(\hat{P}).
\end{equation}
The continuous bath Hamiltonian and the system-bath coupling Hamiltonian are as follows.
\begin{equation}
\hat{H}_b = \int dk^{\prime} \, \epsilon_{k^{\prime}} \hat{c}_{k^{\prime}}^\dagger \hat{c}_{k^{\prime}},
\end{equation}
\begin{equation}
\hat{H}_{c} = \sum_m \int dk^{\prime} \, \left( V_{mk^{\prime}}(\hat{R}) \hat{d}_m^\dagger \hat{c}_{k^{\prime}}
+ V_{mk^{\prime}}^*(\hat{R}) \hat{c}_{k^{\prime}}^\dagger \hat{d}_m \right).
\end{equation}

Our approach begins by discretizing the continuous electronic spectrum of the metal surface,
a common first step in IESH and OSH simulations. To incorporate the electrostatic interactions neglected by this discretization,
we reformulate the problem within an open quantum system framework. Here, the "system" comprises the adsorbate, the discretized
metal surface, and their interaction, while the "bath" is composed of the remaining continuum of electronic states.
Formally, the bath Hamiltonian is $\hat{H}_{b} = \int dk^{'} \epsilon_{k^{'}}c_{k^{'}}^{\dagger}c_{k^{'}}$,
where the integration excludes the discretized wavevectors ($k^{'} \notin k$).
However, in our numerical implementation, the bath is itself represented by a limited number of discrete levels.
This practical approximation should not impact our qualitative findings; therefore,
for the sake of a general discussion, we do not enforce the strict continuum limit for the bath.
For the system-bath coupling, we assume that the molecule-surface interaction is already contained within the system Hamiltonian
through the discretization method, following the work of Tully et al.\cite{tully2009j,miao18,metal-Gardner2023}.

As previously noted, the finite number of discrete energy levels is much smaller than the original continuum,
leading to a small mean energy level spacing $\Delta E$. This spacing is significantly less than the hybridization width $\Gamma$ ($\Delta E < \Gamma$).
Consequently, the timescales for electronic dissipation in the bath ($\tau_b$) and system electron transfer ($\tau_s$) are well separated, with dissipation being orders of magnitude faster (i.e., $\tau_b \ll \tau_s$). This clear separation of timescales makes it reasonable to treat the system-bath interaction within the weak coupling framework.
Under the Born-Markovian approximation, the evolution of density matrix can be described as:

  \begin{equation}
     \Dot{\hat{\rho}}_{s} = -\frac{i}{\hbar}[\hat{H}_{s}, \hat{\rho}_{s}] + \hat{\hat{L}}_{bs}\hat{\rho}_{s}. 
  \end{equation}
The $\hat{\hat{L}}_{bs}$ is the Liouvelle superoperator. It opetrate on the $\hat{\rho}_{s}$ can be given as

\begin{equation}
    \hat{\hat{L}}_{bs}\hat{\rho}_{s} = \frac{1}{\hbar^{2}}\int_{0}^{\infty} d\tau e^{-i\hat{H}t/\hbar} tr_{b}
    ([\hat{H}_{c}(t), [\hat{H}_{c}(t-\tau),e^{i\hat{H}t/\hbar}\hat{\rho}_{s}(t)e^{-i\hat{H}t/\hbar}\bigotimes \hat{\rho}^{eq}_{b} ]])e^{i\hat{H}t/\hbar},
\end{equation}
which is a "Redfield equation" in Schrodinger picture.

The evolution of the single-particle reduced density matrix element, defined as $\hat{\sigma}_{jk} = \text{tr}_{e}\hat{\rho}_{s}\hat{d}^{\dagger}_{j}\hat{d}_{k}$,
is derived by right-multiplying Eq.(5)
with the operator $\hat{d}^{\dagger}_{j}\hat{d}_{k}$ and subsequently performing the trace over the entire electronic Hilbert space.

\begin{equation}\label{eqn:single_body_lvn}
    \Dot{\hat{\sigma}}_{s} = -\frac{\im}{\hbar} \comm{\hat{H}^{orb}_\text{total}}{\hat{\sigma}_{s}} +
    \text{tr}_{e}(\hat{\hat{L}}_{bs}\hat{\rho}_{s}\hat{d}^{\dagger}_{j}\hat{d}_{k}).
\end{equation}
The first two terms in Eq.~\ref{eqn:single_body_lvn} are consistent with our established results (Ref.~\citenum{dou-osh-2025}). However, the derivation of the final term is more involved. It can be rigorously derived by
direct application of Wick's theorem.\cite{Subotnik-2016-jcp-MEEF,Subotnik-2017-jctc-MESH}
Here, we follow the treatment presented in Ref.~\citenum{Subotnik-2016-jcp-MEEF}. Finally, the evolution of the orbital density matrix in the adiabatic representation is governed by:

\begin{equation}\label{eqn:single_body_lvn_ets}
    \Dot{\hat{\sigma}}_{s} = -\frac{\im}{\hbar} \comm{\hat{H}^{orb}_\text{total}}{\hat{\sigma}_{s}}
    -\frac{1}{2\hbar}[\hat{\Gamma}, \hat{\sigma}_{s}]_{+} + \frac{1}{2\hbar}[f(\hat{h}), \hat{\sigma}_{s}]_{+}.
\end{equation}
Here $f$ is the Fermi distribution functions $ f(\epsilon) = \nicefrac{1}{[e^{(\epsilon - \mu_{b})/k_{B}T} + 1]}$ and 
$\hat{H}_{total}^{orb}$ represents the total Hamiltonian in the single-body basis set $\{\ket{j}, j=1, \dots, m\}$:
\begin{equation}\label{eqn:single_body_Hmol}
    \hat{H}^{orb}_\text{total} = \hat{T}_\text{n} +U (\hat{\bm{R}}) + \sum_{jk}^m h_{jk}(\bm{R}) \ketbra{j}{k},  
\end{equation}
$U(\bm{R})$ represents the pure nuclear potential energy.

This equation can be addressed within a quasi-classical framework. To this end, we perform a partial Wigner transformation
of Eq.~\ref{eqn:single_body_lvn_ets} over the nuclear degrees of freedom.
The resulting equation comprises two distinct contributions: 1. The second term constitutes the core of the orbital quantum-classical Liouville equation (o-QCLE).
For this term, we retain the Wigner-Moyal expansion to first order in the nuclear momenta, consistent with the standard approach\cite{QCLE-Kapral1999}.
2. The final term represents electronic dissipation at the metal surface, which occurs on a timescale much shorter than nuclear motion.
This timescale separation justifies the application of the Condon approximation,
allowing us to truncate the corresponding Wigner-Moyal operator to its zero-th order\cite{Nitzan-2015-jpcl-zero-order,Subotnik-2016-jcp-MEEF,Subotnik-2017-jctc-MESH}.
Applying these approximations yields the final working equation:
\begin{equation}\label{eqn:qcle_adiabatic}
\begin{split}
     \frac{\partial \hat{\sigma}^{jk}_{W}(\bm{R},\bm{P},t)}{\partial t} = &-\im \omega^{jk}\sigma^{jk}_{W}(\bm{R},\bm{P},t) + \sum_{l} \sum_{a}\frac{P_{a}}{M_{a}}(\sigma^{jl}_{W}d_{lk}^{a}-d_{jl}^{a}\sigma^{lk}_{W}) \\  
    &- \sum_{a}\frac{P_{a}}{M_{a}} \pdv{\sigma_{W}^{jk}}{R_a} -\frac{1}{2}\sum_{l}\sum_{a} \left( F_{a}^{jl} \pdv{\sigma_{W}^{lk}}{R_{a}} + \pdv{\sigma_{W}^{jl}}{P_{a}} F_{a}^{lk} \right)  + \sum_{a} \pdv{U}{R_a} \pdv{\sigma^{jk}_{W}}{P_a} \\ 
    & -\frac{1}{2}\sum_{l}(\Gamma_{jl}\sigma^{lk}_{W}+\sigma^{jl}_{W}\Gamma_{lk}) + \frac{1}{2}(f(h_{jj})\Gamma_{jk} + \Gamma_{jk}f(h_{kk})),      
\end{split}
\end{equation}
where $\omega^{jk}=(\epsilon_{j}-\epsilon_{k})/\hbar$. The nonadiabatic coupling matrix elements and the force calculated under Hellman Feynman equation are given as: 
\begin{equation}
    d_{jk}^{a}=\bra{\phi_{j}(\bm{R})}\frac{\partial}{\partial R_a}\ket{\phi_{k}(\bm{R})} = -\frac{F^{jk}_{a}}{\hbar\omega^{jk}},
\end{equation}
\begin{equation}
F^{jk}_{a} =  \bra{\phi_{j}(\bm{R})}[\partial h/\partial R_{a}] \ket{\phi_{k}(\bm{R})}.
\end{equation}

To implement the surface hopping approach, the nuclei freedom $\bm{R}$, $\bm{P}$ propagate by the Hamilton equations of motion:
    \begin{gather}
        \dot{\bm{R}} = \frac{\bm{P}}{M}, \\
        \dot{\bm{P}} = -\pdv{U}{\bm{R}} + \sum_{i=1}^{n} \bm{F}_{\lambda_i, \lambda_i}.   
    \end{gather}
    
The nuclei evolve subject to the single-particle forces derived from the occupied orbitals, rather than on a single potential energy surface of an active state.
Concurrently, the electronic subsystem is propagated according to the equation of motion for the single-particle density matrix.
    \begin{equation}
    \begin{split}
         \dot{\sigma}_{ij}(t) = &-\im \omega^{ij} \sigma_{ij}(t) - \frac{\bm{P}}{M} \cdot \sum_k (\bm{d}_{ik} \sigma_{kj} - \sigma_{ik}\bm{d}_{kj}) \\  
        &-\frac{1}{2}\sum_{l}(\Gamma_{il}\sigma_{lj}+\sigma_{jl}\Gamma_{li}) + \frac{1}{2}(f(h_{ii})\Gamma_{ij} + \Gamma_{ij}f(h_{jj})) .\label{Eq:eomo}   
    \end{split}
    \end{equation}
The probability of hopping from the orbital $\lambda_i$ to the orbital $j$ is
    \begin{equation}
        g_{j \gets \lambda_i} = \left\{
        \begin{array}{ll}
            \max \left\{-2\Re(\sigma^{*}_{j \lambda_i}\dot{\bm{R}}\cdot\bm{d}_{j \lambda_i} ) \Delta t'/ \sigma_{\lambda_i \lambda_i}, \ 0 \right\} & j \notin \vec{\lambda} \\
            0 & j \in \vec{\lambda}
        \end{array}
        \right..
    \end{equation}
The stochastic hopping algorithm is identical to that of FSSH, \cite{tully1990} albeit hopping will be considered for each occupied electron.
To minimize hopping, no more than one electron is allowed to transition to a different orbital per nuclear propagating time step. 

The second hopping mechanism originates from the rapid electronic dissipation within the metal surface.
Since the continuous bath degrees of freedom have been traced out, its influence is captured implicitly.
This fast dissipation justifies the use of the secular approximation, wherein we retain only the diagonal
elements of the single-particle density matrix $\sigma$. Consequently, neglecting nuclear dynamics and hopping
events induced by direct non-adiabatic (derivative) coupling,
Eq.~\ref{Eq:eomo} becomes the simplified form:
\begin{equation}
    \dot{\sigma}_{ii} = -\Gamma_{ii}(\sigma_{ii} - f(h_{ii})).
\end{equation}
Our derivation of the hopping probability follows an occupation number formalism, where each orbital is either occupied (1) or unoccupied (0), analogous to the treatment in Ref.~\citenum{Subotnik-2015-jcp-mbah}. The specific probability for a hop to occur involving orbital $i$ is:

\begin{equation}
        g_{i' \gets i} = \left\{
        \begin{array}{ll}
            \max \left\{\Gamma_{ii}(1-f(h_{ii})), \ 0 \right\} & (i' \in 0, i \in 1)\\
            \max \left\{\Gamma_{ii}f(h_{ii}), \ 0 \right\} & (i' \in 1, i \in 0)
        \end{array}
        \right..
    \end{equation}
This hopping probability is governed by the hybridization function, $\Gamma_{ii}$, which quantifies the spectral broadening
and dictates the inverse lifetime of the electronic state. Our model employs the wide-band and Condon approximations,
justified by the separation of timescales between rapid electronic dissipation at the metal surface and slower nuclear motion.
This allows $\Gamma_{ii}$ to be treated as an energy-independent constant. We assign it a value comparable to the characteristic energy scale of the problem—specifically, the level spacing near the Fermi energy, 
where electronic excitations are dominant.

For comparative analysis, we have implemented Tully's electronic thermostat method\cite{tully2012fd} within our OSH framework.
In this approach, electron dynamics are propagated using the first three terms of Eq.~\ref{Eq:eomo}. The second hopping mechanism
is replaced by a stochastic process: an occupied orbital $i$ and an unoccupied orbital $j$ are randomly selected.
A hopping event from $i$ to $j$ is then accepted with a probability: min(1, $e^{-(h_{jj} - h_{ii}) / k_B T}$).
This entire procedure is triggered at each time step with a probability of $\Delta t /\tau$, where $\tau$ is the characteristic timescale of the thermostat.
A comprehensive discussion and comparison of the results are presented subsequently.

\section{Results and discussion}\label{sec:results}

To test our new electron thermostat method, we employ the Newns-Anderson model, which has been successfully applied
to study non-adiabatic effects at gas-metal interfaces.\cite{tully2009j,tully2009s,ouyang16,miao18,amber2022}
To validate the accuracy of our method, we used the highly accurate but computationally expensive Hierarchical
Equations of Motion (HEOM) method\cite{heom-1989} as a benchmark. To make these benchmark calculations feasible,
minor adjustments to the system parameters were necessary. For larger systems where HEOM becomes prohibitively expensive,
our comparison is confined to assessing the performance of our OSH method with and without the electronic thermostat.

\subsection{Simulation details}
The Newns-Anderson Hamiltonian with one impurity molecule orbital and a continuum of metal electron orbitals that discretized 
into M orbitals is given as:

\begin{equation}\label{eqn:newns_anderson}
\begin{split}
    \hat{H}_\text{el}(\bm{R}) =  &U_{0}(\bm{R}) + (U_{1}(\bm{R})-U_{0}(\bm{R}))c^{\dagger}_{a}c_{a}+ \sum_{n=1}^{M}\epsilon_{n} c^{\dagger}_{n}c_{n} \\ 
    &+\sum_{n=1}^{M} (V_{n}(\bm{R}) c^{\dagger}_{a}c_{n}+V_{n}(\bm{R})^{*} c^{\dagger}_{n}c_{a}),
\end{split}    
\end{equation}
where $c^{\dagger}_{\epsilon} (c_{\epsilon})$ and $c^{\dagger}_{a} (c_{a})$ are the fermionic creation (annihilation)
operators of metal and molecule orbital, respectively. $V(\bm{R})$ are the couplings between molecular and metal orbitals. 
$U_{1}(\bm{R}), U_{0}(\bm{R})$ are the potential energy surfaces of the anionic and neutral molecule, respectively.
We consider a 1D double well potential following Refs.~\cite{ouyang16,miao18,amber2022,metal-Gardner2023}

\begin{equation}
    U_0(R) = \frac{1}{2} M \omega^{2} R^{2}, 
\end{equation}
\begin{equation}
   U_1(R) - U_0(R) = -M \omega^2 g R + \frac{1}{2} M \omega^2 g^2 + \Delta G.
\end{equation}
Here, the displacement parameter $g$ is related to reorganization energy by $E_\text{r} = \frac{1}{2} M \omega^2 g^2.$
The parameters used for our model are (units: a.u.) $\text{m} = 2.0 \times 10^{3}$, $\omega = 2.0 \times 10^{-4}$, $g = 20.6097$,
$\Delta G_{0} = -3.8 \times 10^{-3}$ and $kT = 9.5 \times 10^{-4}$. These parameters are prepared according to Ref.~\citenum{ouyang16,miao18}.
The wide-band spectral density function is used and the hybridization function,
\begin{equation}   
    \Gamma(\epsilon; \bm{R}) \equiv 2 \pi \sum_k \abs{V(\epsilon, \bm{R})}^2 \delta(\epsilon - \epsilon_k),
\end{equation}
is assumed to be a constant $\Gamma$ across the interval $E_{-} = -W / 2 + \mu$ and $E_{+} = W / 2 + \mu$.
Here, $W$ and $\mu$ are the bandwidth and the chemical potential, respectively. 
The trapezoid discretization procedure is adopted following Ref.~\citenum{miao18},
as it is reported to be numerically more stable\cite{miao18} and equally accurate as Gauss-Legendre procedure 
for small bandwidths.\cite{metal-Gardner2023} Throughout this section, the occupation of the metal orbitals are prepared at "zero temperature", 
i.e., all the diabatic orbitals below Fermi level $\mu$ are occupied. The impurity molecular orbital, on the other hand, is unoccupied.

To evaluate the performance of electron thermostat, we initialized an ensemble of trajectories from the Boltzmann distribution of $U_0$,
imparting additional initial kinetic energy.\cite{wigner_function_Sun2010,miao18} The dynamics were compared under two scenarios:
(1) with only the electronic thermostat, and (2) with the electronic thermostat plus an external nuclear friction force.
This friction, parameterized universally by $\gamma_\text{ext} = 2 \omega$, models the contribution from phononic energy dissipation into the surface.

\subsection{Benchmarked with HEOM methods}
The simulated time-dependent kinetic energy and impurity hole population are compared in Fig. 1 across three methods:
HEOM (benchmark), OSH, and OSH with electron thermostat (labeled OSH-ETS).
All HEOM simulations were performed with the HierarchicalEOM.jl package.\cite{HierarchicalEOM-jl2023,QuantumToolbox-jl2025}
The HierarchicalEOM.jl software package provides a user-friendly and efficient tool for simulating complex open quantum systems,
including non-Markovian effects due to non-perturbative interaction with one, or multiple, environments.
The initial total energy was set to 2kT, resulting in an initial kinetic energy of kT.
A displacement parameter of g = 15.7706 was used to enhance computational efficiency.
In our simulations, the hybridization function is represented by a Lorentzian form with $\Gamma = 6.4 \times 10^{-4}$.
The Fermi energy is set to zero. The Fermi distribution is approximated using a Padé expansion,
and the HEOM method is executed with 4 exponential terms and a hierarchy tier of 4 to ensure converged results.
The electronic bandwidth is set to $W = 6.4 \times 10^{-3}$, and the continuum is represented by 40 discrete electron levels.

\begin{figure}[htbp]
    \centering
    \includegraphics[width=0.88\linewidth]{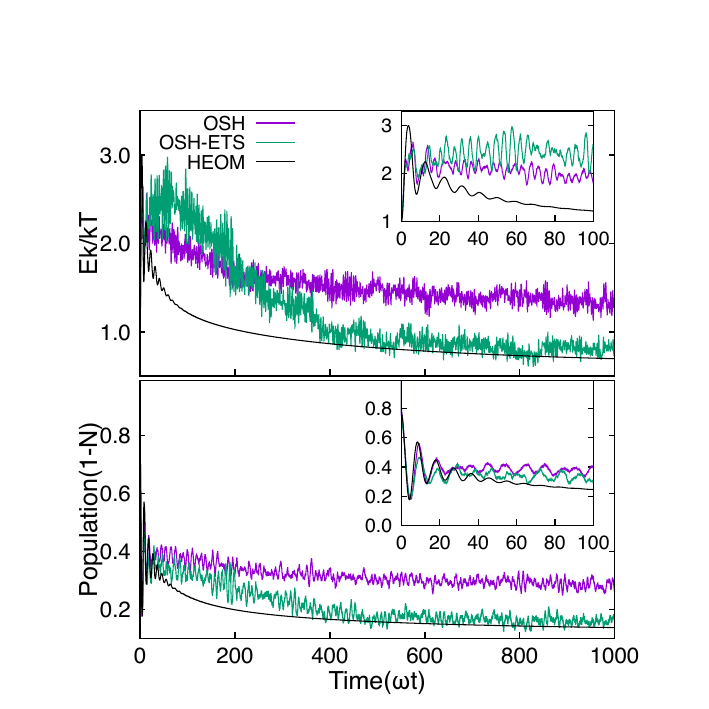}
    \caption{Time evolution of the impurity hole population and kinetic energy, comparing the standard OSH,
    OSH with electronic thermostat (OSH-ETS), and HEOM methods. The inset shows the short-time evolution of the same quantities.} 
\end{figure}

As shown in Fig. 1, the OSH method with our electronic thermostat (OSH-ETS) successfully drives both the population
and kinetic energy to converge with the HEOM benchmark results at long times. In contrast, while the standard OSH method (without a thermostat)
reaches an equilibrium more rapidly, this equilibrium state exhibits a significant discrepancy from the HEOM result.
A detailed examination of the dynamics reveals that OSH and OSH-ETS yield similar trajectories for the first 20 $\omega t$.
Beyond this point, the electronic thermostat in OSH-ETS becomes active, guiding the impurity hole population and kinetic energy
toward the correct thermal equilibrium. This successful reproduction of the benchmark dynamics demonstrates the accuracy
and utility of our new electronic thermostat method.

\subsection{Comparison under pure electron thermostat}

To evaluate the performance of our new thermostat method, we simulated a wide range of molecule-metal coupling strengths (Fig. 2).
We use the bandwidth $W$ and number of states $M$ from Ref.~\citenum{miao18}, where convergence of the IESH method was demonstrated.
Since our previous work established that OSH converges at least as fast as IESH with respect to the state number at a fixed $W$\cite{dou-osh-2025},
we retain these parameters for all calculations.

The results demonstrate that only the methods incorporating an electronic thermostat—both our OSH-ETS and Tully's approach (labled OSH-$\tau$ = 100)—faithfully
reproduce the correct equilibrium kinetic energy and impurity hole population. In contrast, the standard OSH method without a thermostat
shows significant deviations in the impurity population across different coupling regimes, underscoring the critical role of
the electronic dissipation channel. The simulations lead to two principal conclusions: Essential Role of Electronic Thermostat:
Accurate equilibrium kinetic energy and impurity population are only achieved when an electronic thermostat (either OSH-ETS or Tully's method)
is employed. Failure of Standard OSH: The standard OSH method, lacking this dissipation mechanism, gives rise to significant errors
in the impurity population that persist across different coupling strengths.

\begin{figure}[htbp]
    \centering
    \includegraphics[width=0.98\linewidth]{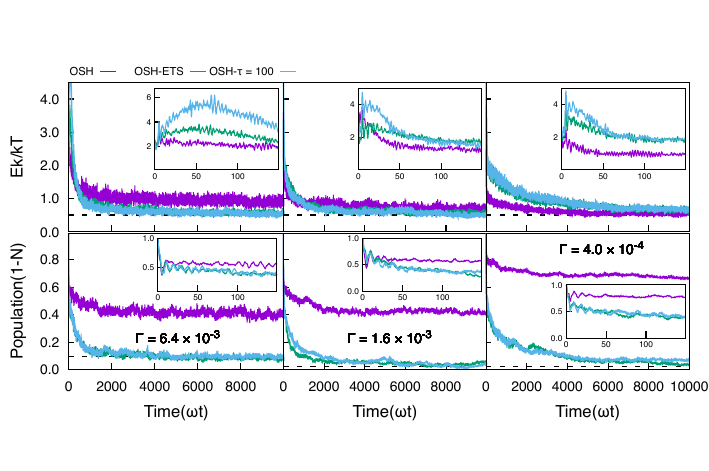}
    \caption{Kinetic energy (top) and impurity hole population dynamics (bottom) as a function of time.
    Results are shown for three methods: OSH, OSH-ETS, and OSH with Tully's electronic thermostat approach ($\tau$ = 100),
    across a range of coupling strengths: (left) $\Gamma = 6.4 \times 10^{-3}$, (middle) $\Gamma = 1.6 \times 10^{-3}$,
    and (right) $\Gamma = 4.0 \times 10^{-4}$. The insets zoom in on the short-time behavior.} 
\end{figure}

\subsection{Comparison with external friction }
Figure 3 displays the impurity hole population dynamics in the presence of external nuclear friction across various coupling strengths.
A key observation is that the inclusion of nuclear friction drives all methods toward a similar equilibrium population,
with the electronic thermostats primarily serving to accelerate the equilibration process.
In most cases, our method and Tully's yield nearly identical results. A significant deviation,
however, occurs at the specific coupling strength of $\Gamma = 6.4 \times 10^{-3}$. At this value, Tully's method produces a result similar
to the case without any electronic thermostat, whereas our OSH-ETS method maintains an accurate population.
To verify the robustness of our finding, we conducted additional simulations at this coupling with different friction coefficients
$\gamma$ (see Figure S1, Appendix). These tests confirm that while varying $\gamma$ alters the kinetic rates, it does not affect the long-time population limit.
Under this specific condition, our method integrated with the OSH framework provides a more accurate result than Tully's.
For Tully's method, the outcome was found to be largely insensitive to the value of the thermostat timescale $\tau$.

\begin{figure}[htbp]
    \centering
    \includegraphics[width=0.8\linewidth]{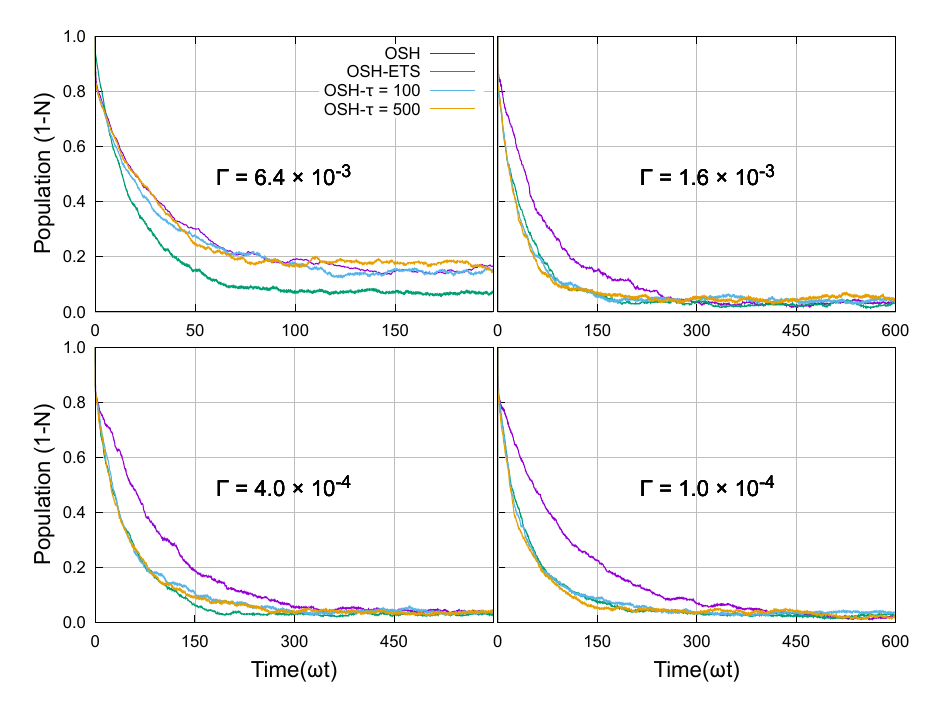}
    \caption{Impurity hole population dynamics as a function of time for OSH, OSH-ETS and OSH ($\tau =$ 100 and 500 for Tully's electronic thermostat method).} 
    \label{fig:pop_eq_methods_compare}
\end{figure}

\section{Conclusion}
In summary, we have developed a novel electronic thermostat method to model energy dissipation into a continuum of surface states.
This approach is grounded in open quantum system theory and seamlessly integrates with our previously established OSH framework.
The method was rigorously validated against the numerically exact HEOM approach, demonstrating that OSH with our thermostat correctly
drives the system toward the accurate equilibrium kinetic energy and population.

Extensive comparisons across a broad parameter range reveal that,
in the presence of a pure electronic thermostat, both our method and Tully's achieve accurate equilibrium states. When external nuclear friction is included,
the performance of our method is largely comparable to Tully's electron thermostat method.

Our analysis demonstrates that both our method and Tully's electronic thermostat are capable and efficient for simulating metal surfaces,
delivering comparable results. The choice between them can be guided by the specific problem: Tully's method, which conserves electron number,
remains a robust option for closed systems like Auger recombination. Our method, derived from open system theory, offers a complementary
and formally rigorous pathway for handling energy dissipation, integrating seamlessly with our broader OSH framework.  

\begin{acknowledgement}

W. D. thanked the funding from the National Natural Science Foundation of
China (no. 22361142829) and Zhejiang Provincial Natural Science Foundation (no. XHD24B0301)). 
Y.-T. Ma thanked the project (Grant No. ZR2021MB081) supported by Shandong Provincial
Natural Science Foundation and thanked Dr. Yu Wang and Mr. Ruihao Bi for helpful discussion.
The authors thanked the Westlake University Supercomputer Center for facility support and technical assistance. 

\end{acknowledgement}


\appendix



\section{Compare parameters for external fiction}

At a coupling strength of $\Gamma = 6.4 \times 10^{-3}$, our electron thermostat method shows a marked deviation from Tully's result (Fig. 3). To establish robustness, we systematically varied the friction coefficient $\gamma$ at this coupling.
As shown in Fig. S1, while $\gamma$ modulates the transient kinetics, it does not affect the steady-state population.
This finding underscores that, under this specific conditions,
our OSH-electron thermostat method achieves a more physically accurate representation than Tully's.

\begin{figure}[htbp]
    \centering
    \includegraphics[width=0.7\linewidth]{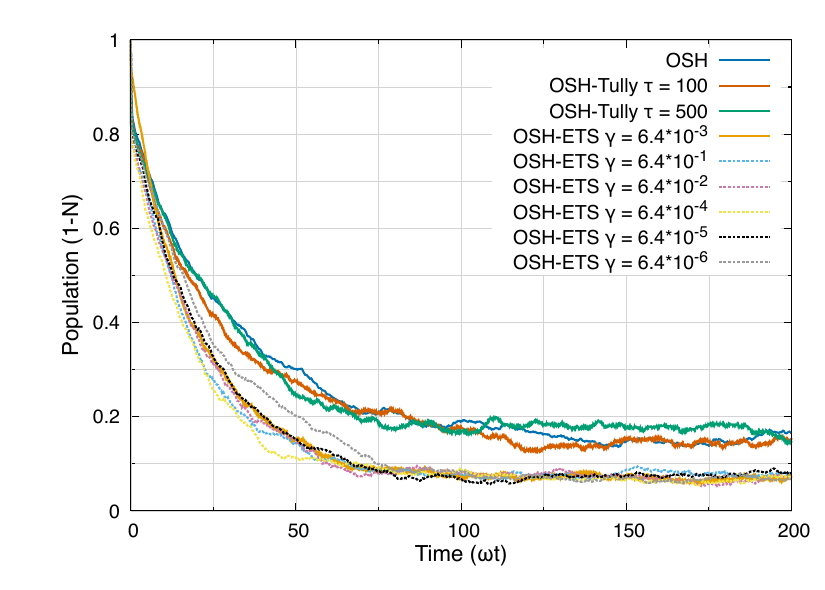}
    \caption{Impurity hole population dynamics and kinetic energy as a function of time for OSH, OSH-ETS and 
    OSH with Tully's electronic thermostat approach ($\tau$ = 100 and 500).} 
\end{figure}

\bibliography{achemso-demo}

\end{document}